\documentclass[twocolumn,showpacs,superscriptaddress,preprintnumbers,amsmath,amssymb]{revtex4-1}

\usepackage{graphicx}% Include figure files
\usepackage{dcolumn}% Align table columns on decimal point
\usepackage{appendix}
\usepackage{bm}% bold math

\usepackage[usenames,dvipsnames]{xcolor}
\usepackage{subfigure}
\usepackage{bbm}
\usepackage{enumerate}
\usepackage[
  bookmarks=true,
  colorlinks,
  linkcolor=black,
  urlcolor=black,
  citecolor=black,
  plainpages=false,
  pdfpagelabels,
  final,
  breaklinks=true
]{hyperref}

\usepackage{titlesec}
\usepackage{array}
\usepackage{dsfont}
\usepackage{physics}
\usepackage[normalem]{ulem}

\begin{document}

\title{Metrological robustness of high photon number optical cat states}

\author{Philipp Stammer}
\email{philipp.stammer@icfo.eu}
\affiliation{ICFO - Institut de Ciències Fotòniques, The Barcelona Institute of Science and Technology, 08860 Castelldefels (Barcelona), Spain}
\affiliation{Atominstitut, Technische Universität Wien, Stadionallee 2, 1020 Vienna, Austria}

\author{Tomás Fernández Martos}
\affiliation{ICFO - Institut de Ciències Fotòniques, The Barcelona Institute of Science and Technology, 08860 Castelldefels (Barcelona), Spain}

\author{Maciej Lewenstein}
\affiliation{ICFO - Institut de Ciències Fotòniques, The Barcelona Institute of Science and Technology, 08860 Castelldefels (Barcelona), Spain}
\affiliation{ICREA, Pg.\ Lluis Companys 23, 08010 Barcelona, Spain}

\author{Grzegorz Rajchel-Mieldzioć}
\email{grajchel@icfo.eu}
\affiliation{ICFO - Institut de Ciències Fotòniques, The Barcelona Institute of Science and Technology, 08860 Castelldefels (Barcelona), Spain}
\affiliation{NASK National Research Institute, ul. Kolska 12 01-045 Warszawa, Poland}

\date{\today}

\begin{abstract}

In the domain of quantum metrology, cat states have demonstrated their utility despite their inherent fragility with respect to losses. 
Here, we introduce noise robust optical cat states which exhibit a metrological robustness for phase estimation in the regime of high photon numbers.
These cat states are obtained from the intense laser driven process of high harmonic generation (HHG), and show a resilience against photon losses. 
Focusing on a realistic scenario including experimental imperfections we opt for the case in which we can maximize the lower bound of the quantum Fisher information (QFI) instead of analyzing the best case scenario. 
We show that the decrease of the QFI in the lossy case is suppressed for the HHG-cat state compared to the even and odd counterparts. 
In the regime of small losses of just a single photon, the HHG-cat state remains almost pure while the even/odd cat state counterparts rapidly decohere to the maximally mixed state.
More importantly, this translates to a significantly enhanced robustness for the HHG-cat against photon loss, demonstrating that high photon number optical cat states can indeed be used for metrological applications even in the presence of losses.

\end{abstract}

\maketitle
%\tableofcontents
%\begin{multicols}{2}

% \emph{Introduction.} -- 
\section{Introduction}
Quantum metrology exploits properties such as entanglement or squeezing to achieve highly sensitive measurements of physical parameters with potentially higher precision than can be achieved using purely classical approaches~\cite{giovannetti2004quantum,giovannetti2011advances}. A standard way of quantifying the metrological limits for parameter estimation is the use of the quantum Fisher information (QFI)~\cite{liu2020quantum}. The QFI sets an upper bound on the achievable information gain from any measurement of a quantum state $\hat{\rho}(\bm{\theta})$, with encoded unknown parameters $\bm{\theta}$, under the optimal choice of observables. This is directly tied to a lower bound on the achievable precision of the measurement, given by the quantum Cramér-Rao bound~\cite{helstrom1969quantum}. 
The QFI gives a measure of the sensitivity of the quantum state $\hat{\rho}(\bm{\theta})$ to a change of a certain parameter $\bm{\theta}$; by maximizing this quantity, upper bounds on the sensitivity for specific measurements can be found. However, for any realistic setting one needs to take into account the effect of noise, which limits the attainable precision of measurements~\cite{toth2014quantum, meyer2021fisher}. 

One of the possibilities for surpassing classical limits of metrology is to use quantum superpositions of coherent states, known as Schrödinger cat states~\cite{gerry1997quantum,gilchrist2004schrodinger, ralph2003quantum}, which can be advantageous for phase estimation~\cite{joo2011quantum, lee2015quantum}. 
Cat states are implemented in various systems, e.g., employing one-axis twisting dynamics~\cite{Kitagawa_1993,Wineland_1994}, which can be implemented with ultra-cold atoms in optical lattices~\cite{Plodzien2020, Plodzien2022, Plodzien2023generation, Hernandez2022,Dziurawiec2023,yanes2023spin}.
Here, we focus on optical cat states, conventionally given by the superposition of two coherent states of opposite phase, the most archetypal being the even/odd cat states~\cite{gerry1997quantum}. The coherent states in the superposition are macroscopically distinguishable for high values of the mean photon number, i.e. coherent state amplitude. 
In particular, for phase estimation the measurement uncertainty is typically inversely proportional to the average photon number~\cite{escher2011general}, which makes high photon number optical cat states a desirable resource in quantum metrology.
However, conventional approaches for generating optical cat states are limited by their low photon number~\cite{ourjoumtsev2006generating, ourjoumtsev2007generation}. In addition, cat states are very vulnerable to the decoherence introduced by photon loss, which becomes severe very rapidly with increasing photon number~\cite{glancy2008methods, zhang2013quantum}. Nonetheless, optical cat states are still used in quantum metrology since they can provide an enhanced phase sensitivity~\cite{yu2018maximal}. 

In addition to the conventional approaches for generating optical cat states~\cite{ourjoumtsev2007generation, glancy2008methods}, there have been recent advances in the generation of high photon number optical cat states, by using the platform of intense laser matter interaction for quantum state engineering of light~\cite{lewenstein2021generation, stammer2023quantum, bhattacharya2023strong}. In particular, the processes of high harmonic generation (HHG)~\cite{lewenstein2021generation, rivera2022strong} and above-threshold ionization~\cite{rivera2022light} have been utilized to generate non-classical states of light, especially massive entangled states~\cite{stammer2022high, stammer2022theory, stammer2023entanglement} and high photon number optical cat states~\cite{stammer2022high, rivera2022strong, rivera2021new}. The cat state obtained from HHG is given by the superposition of two coherent states with amplitude $\alpha$, which have a relative amplitude displacement of $\delta \alpha$. By tuning the relative displacement $\delta\alpha$, the macroscopic distinguishability between the two coherent states can be controlled. In theory, the mean photon number of the HHG-cat state can be made arbitrarily large, by increasing $\alpha$, without affecting its resilience against decoherence (which depends on $\delta \alpha$). This makes it potentially more suitable for quantum metrology applications in realistic scenarios including noise than conventional cat states. In practice, the generation of an optical HHG-cat state with high average photon number of $\expval*{\hat{N}}\approx 150$ has been shown~\cite{lamprou2023nonlinear}, orders of magnitude higher than those of conventional schemes. 

In this work, we explore the usefulness of the optical cat state obtained from HHG for metrological applications in a homodyne phase sensing configuration, and we show its resilience against photon loss in the high photon number regime. 
This ultimately allows to make use of the features arising from the quantum superposition in the high photon number regime. 
In particular, we do not aim to focus on the best case scenario, which possibly does not occur in realistic setups, but rather consider the setting in which experimental uncertainties and losses occur.

% \emph{Noise robust optical cat states.} -- 
\section{Noise robust optical cat states}

\subsection{Optical cat states from HHG}
The aforementioned HHG-cat state is obtained when driving atoms by an intense laser field $\ket{\alpha}$ and performing conditioning approaches on the process of HHG~\cite{lewenstein2021generation}. 
These conditioning measurements, extensively described in~\cite{stammer2022high, stammer2022theory, stammer2023quantum}, lead to a quantum operation acting on the driving field mode and result in the generation of a high photon number optical cat state. It has been shown by tomographic reconstruction~\cite{lewenstein2021generation, rivera2022strong} of the state that it has the following form
\begin{equation}
\label{eq:cat_HHG}
    \ket{\psi_{\mathrm{HHG}}} = \frac{1}{\sqrt{N_{\mathrm{HHG}}}}\left(\ket{\alpha + \delta \alpha} -    \xi\ket{\alpha}\right),    
\end{equation}
where $\xi = \bra{\alpha}\ket{\alpha+\delta\alpha}$ and $N_{\mathrm{HHG}} = 1-\abs{\xi}^2$ is the normalization. This approach can be used for quantum state engineering of light~\cite{stammer2023quantum, stammer2023role, stammer2023entanglement} in different target systems~\cite{rivera2022quantum, rivera2023bloch} or when considering driving the processes with non-classical states of light~\cite{gorlach2023high, stammer2023limitations, pizzi2023light}. One of the key factors that allows for achieving high photon numbers using this approach is the fact that the generated superpositions come from the back-action of the laser-matter interaction processes on the initial coherent state of the driving laser field~\cite{lewenstein2021generation}. This intense laser field inherently contains a very high number of photons, which will be used for the generation of high photon number non-classical field states. In the case of HHG, the generation of high-energy harmonic photons comes at the expense of a depletion of the original laser field, which manifests itself in the amplitude displacement $\delta\alpha$, and conditional measurements allow to generate the superposition in Eq.~\eqref{eq:cat_HHG}. This displacement can be controlled experimentally by altering the laser-atom interaction conditions~\cite{stammer2023quantum}, which can be used to engineer the quantum superposition, such as altering the distinguishability of the two states.
The prefactor $\xi$ of the second term determines the distinguishability of the two states in the superposition, which can not be made arbitrarily large. However, in the following we will see that it is exactly this feature which makes the HHG-cat much more robust against losses than the conventional even/odd cat states. 
This is because the even $(+)$ and odd $(-)$ cat state are given by an equal superposition of coherent states with the same amplitude but opposite phase
\begin{align}
\label{eq:cat_conventional}
    \ket{\psi_\pm}  = \frac{1}{\sqrt{N_{\pm}}}(\ket{\alpha} \pm \ket{-\alpha}),
\end{align}
where $N_{\pm}$ is the normalization factor~\cite{gerry1997quantum}. Comparing the even/odd cat state with the HHG-cat in \eqref{eq:cat_HHG} we can see that the average photon number can only be varied by controlling the amplitude $\alpha$. This also affects the distinguishability of the two states in the superposition with an exponentially decreasing overlap for increasing $\alpha$. In the following we will see that this makes the even/odd cat states vulnerable to photon losses in the high-photon number regime.

\subsection{\label{sec:noisy_cats}Noise robustness}

Typically, the primary source of decoherence in optical experiments comes from photon losses, which can be modeled by introducing a fictitious beam-spliter (BS) with transmission efficiency $\eta$ and tracing out the reflected part (index $r$) of the output~\cite{glancy2008methods,zhang2021quantifying,rohde2007practical}. The transmitted part (index t) is then expressed as
\begin{equation}
\hat{\rho}_{ \mathrm{t}}^{\eta} = \Tr_{\mathrm{r}}\big[\dyad*{\Psi^{\eta}}\big],
    \label{eq:noisemodel}
\end{equation}
where $\ket*{\Psi^{\eta}} = \hat{U}_{\mathrm{BS}}\big[\!\arccos{(\sqrt{\eta})}\big]\ket{\psi}_0\ket{0}_1$, for the state $\ket{\psi}_0$ and the vacuum $\ket{0}_1$ in the input modes of the BS. 
For the even/odd-cat states we have the mixture
\begin{equation}
    \begin{split}
        \hat{\rho}_{\pm}^{\eta} = %& \frac{1}{N_{\pm}}(\ket{\alpha\sqrt{\eta}}\bra{\alpha\sqrt{\eta}} \pm \gamma\ket{\alpha\sqrt{\eta}}\bra{-\alpha\sqrt{\eta}} \\
        %& \pm \gamma\ket{-\alpha\sqrt{\eta}}\bra{\alpha\sqrt{\eta}} + \ket{-\alpha\sqrt{\eta}}\bra{-\alpha\sqrt{\eta}}), \\ 
        (1 - \zeta_{\pm}^{\eta})\dyad*{\psi^{\eta}_{\pm}} + \zeta_{\pm}^{\eta}\dyad*{\psi^{\eta}_{\mp}},
    \end{split}
\label{eq:lossyevenodd}
\end{equation}
where $\ket*{\psi^{\eta}_{\pm}} = (N_{\pm}^{\eta})^{-1/2}\big(\ket*{\alpha\sqrt{\eta}} \pm \ket*{-\alpha\sqrt{\eta}}\big)$ are the normalized even/odd cat states with a reduced amplitude by the factor $\sqrt{\eta}$.
Note that Eq.~\eqref{eq:lossyevenodd} shows that an initial pure even/odd cat state reduces to an equal mixture of even and odd cat states when $\zeta_\pm^\eta = 1/2$, and is given by
\begin{align}
    \zeta_{\pm}^{\eta} = \frac{N_{\mp}^{\eta} }{ 2N_{\pm}}\left( 1 - e^{-2|\alpha|^2(1-\eta)} \right).
\end{align}

Due to the exponential we can see that for cat states of large amplitude the state becomes maximally mixed already for small losses.
It is also this parameter that determines the purity of $\hat{\rho}_{\pm}^{\eta}$, which is given by
\begin{align}
    \gamma_{\pm}^{\eta} \equiv \Tr[(\hat{\rho}_{\pm}^{\eta})^2]  = \left(\zeta_{\pm}^{\eta} \right)^2 + \left(1-\zeta_{\pm}^{\eta} \right)^2,
\end{align}
and rapidly decreases for increasing photon number even in the presence of small losses~\cite{serafini2004minimum}. 
Therefore, for any realistic scenario the even/odd cat states will rapidly decohere into a 50/50 mixture of even and odd cat states with the reduced amplitude $\alpha\sqrt{\eta}$ due to losses. 

Let us now consider the pure HHG-cat state in Eq.~\eqref{eq:cat_HHG} and its exposure to photon losses such that the resulting mixed state in its spectral decomposition after the noise channel reads (the exact expressions are given in Appendix \ref{sec:appendix_HHG})
\begin{equation}
    \hat{\rho}_{\mathrm{HHG}}^{\eta} = \lambda^{\eta}_{+}\dyad*{\lambda^{\eta}_{+}} + \lambda^{\eta}_{-}\dyad*{\lambda^{\eta}_{-}}.
    \label{eq:lossyHHGcat}
\end{equation}

In this case, the purity of the mixture is given by
\begin{align}
\label{eq:purity_HHG}
    \gamma_{\mathrm{HHG}}^{\eta} = \left(\lambda^{\eta}_{+} \right)^2 + \left(\lambda^{\eta}_{-} \right)^2.    
\end{align}

\begin{figure}[ht]
    \centering
    \includegraphics[width=1\columnwidth]{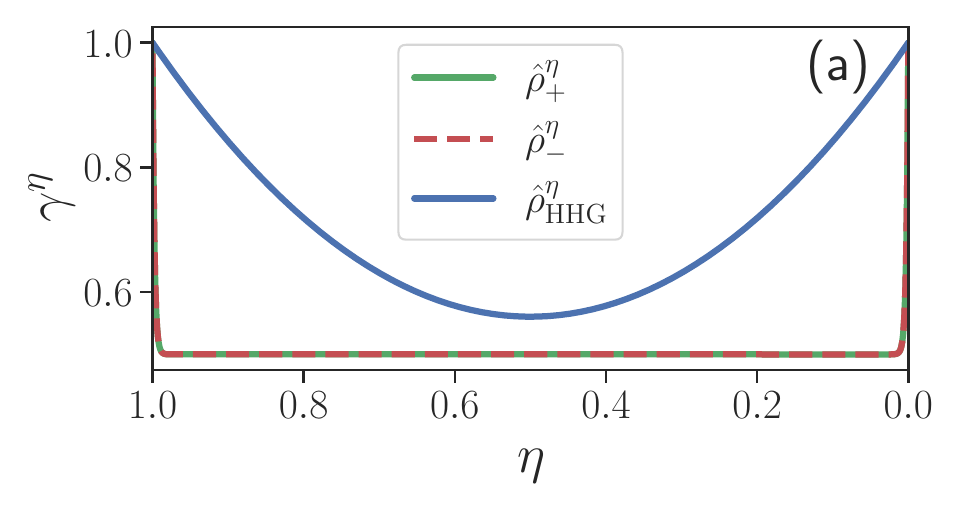}
    \includegraphics[width=1\columnwidth]{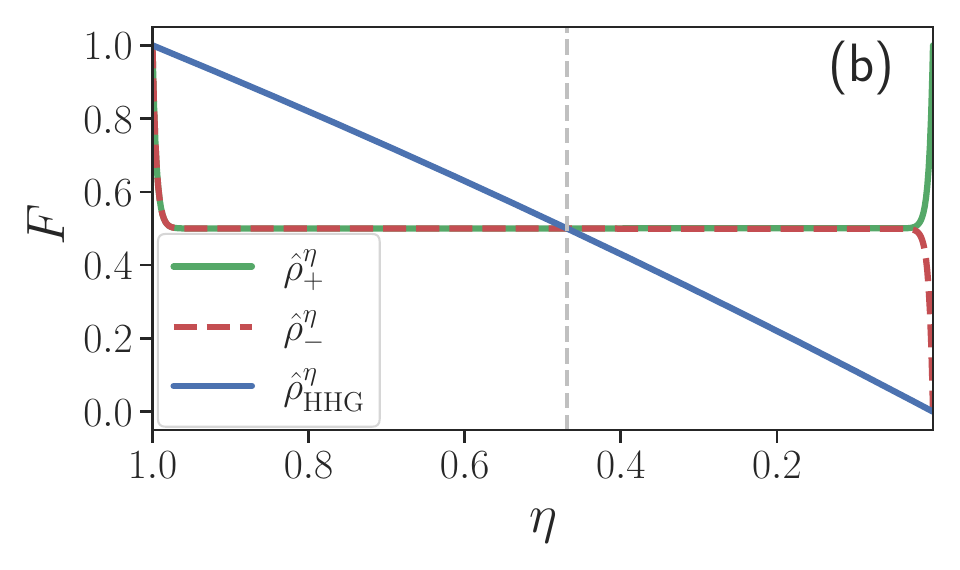}
    \caption{(a): Purity $\gamma^{\eta}$ of the noisy cat states $\hat{\rho}^{\eta}$. (b): Fidelity between the noisy cat states and their pure counterparts. The pure states have a reduced amplitude by a factor of $\sqrt{\eta}$ to maximize the fidelity. A vertical dashed line at $\eta = 0.469$ (53.1\% photon loss) indicates up to which $F_{\mathrm{HHG}} > F_{\pm}$. Note that for $\eta \to 0$ all the noisy states become $\dyad{0}$. 
    In both plots, we consider an initial average photon number of $\expval*{\hat{N}}=100$ before the noise channel, having $\alpha = 10$ for the even/odd cat states, while $\alpha' = 10.5$ and $\delta\alpha = -0.5$ for the HHG-cat. }
    \label{fig:fidelitynoisy}
\end{figure}

Finally, we show the purity of the different cat states as a function of $\eta$ in Fig.~\ref{fig:fidelitynoisy} (a) for an initial photon number of $\langle \hat{N}\rangle = 100$.
We clearly see the rapid decrease of purity to the maximally mixed state with $\gamma_{\pm}^{\eta} = 1/2$ for the even/odd cat state in this photon number regime already for very small losses~\cite{serafini2004minimum}. 
In contrast, the HHG-cat state exhibits only very little mixedness in this regime of small losses.
Furthermore, the HHG-cat state never becomes maximally mixed, i.e. $\gamma_{\mathrm{HHG}}^{\eta} > 1/2$ for the 2-dimensional decomposition, and outperforms the even/odd cat state for all $\eta$.  
Notably, the purity of the HHG-cat is not affected by the average photon number since the eigenvalues in~\eqref{eq:lossyHHGeigenvalues} only depend on the relative displacement $\delta \alpha$, while in contrast, the even/odd cat states asymptotically tend to a constant $\gamma_{\pm}^{\eta} = 1/2$ for increasing $|\alpha|$. This further manifests the robustness of the HHG-cat state in the high-photon number regime since increasing the mean photon number does not imply larger decoherence.

Besides the purity, which measures how mixed the noisy states become due to losses, we shall quantify the state disturbance of the pure cat states subjected to the noise channel. For this, we use the fidelity between the pure and the noisy states $F(\dyad{\psi}, \hat{\rho}^{\eta}) = \bra{\psi}\hat{\rho}^{\eta}\ket{\psi}$ to quantify how close the two states are~\cite{jozsa1994fidelity}. 
In Fig.~\ref{fig:fidelitynoisy} (b) we show the fidelity between the noisy cat states with their pure counterparts for an initial mean photon number of $\expval*{\hat{N}}=100$. 
Note that for each $\eta$ we consider the pure state to be reduced in amplitude by a factor of $\sqrt{\eta}$, in order to maximize the fidelity with the noisy state which has a reduced amplitude due to the losses. 

One of the key result is that the HHG-cat state is significantly more robust than the even/odd cat states for a wide range of up to 53.1\% photon loss.
We further emphasize, that the advantage of the HHG-cat state robustness is most pronounced in the important regime of small photon losses, in which the even/odd cat state counterparts exhibit significantly reduced purity and fidelity and rapidly decay to the 50/50 mixture. In particular, for the $\langle \hat{N} \rangle = 100$ photon states considered in Fig.~\ref{fig:fidelitynoisy}, the even/odd cat states are well approximated by the maximally mixed state $\gamma_\pm^\eta = 1/2$ already for $\eta = 0.983$, while the HHG-cat nearly remains a pure state with $\gamma^\eta_{\mathrm{HHG}} \approx 0.97$.
In addition, the fidelity of the even/odd cat states reduces to $F_\pm \approx 1/2 $ for $\eta = 0.965$, while the noisy HHG-cat is close to its pure counterpart with $F_{\mathrm{HHG}} \approx 0.97$.
To further emphasize the robustness of the HHG-cat for small photon losses, we show the derivative of the purity $\gamma^\eta$ and of the Fidelity $F(\eta)$ with respect to $\eta$ at $\eta = 1$ (see Fig.\ref{fig:fidelity_derivative} in the Appendix \ref{sec:appendix_derivative}).
While the even/odd cat state exhibit a significant decrease in purity and Fidelity for very small photon loss, which increases for increasing photon number. In contrast, the HHG-cat is robust against small photon losses across the entire range of photon numbers. 

In summary, the HHG-cat state is significantly more robust for high photon numbers due to its unique structure. It is the relative displacement $\delta \alpha$ in Eq.~\eqref{eq:cat_HHG} which determines the distinguishability between the two states in the HHG-cat, which remains constant even if the HHG-cat is growing in size by means of the average photon number (due to increasing $\alpha$).
In contrast, for the even/odd cat states an increasing displacement $\alpha$ leads to an increased distinguishability, which causes a rapid decoherence for high photon numbers already for very small losses of just a few photons. 
This ultimately makes the even/odd cat states much more vulnerable compared to the robust HHG-cat state. The HHG-cat state is protected against the losses due to the indistinguishability of the two states in the superposition, while still being of macroscopic size.

% \emph{Metrological robustness.} -- 
\section{Metrologically robust cat states}

To demonstrate the metrological robustness of the optical cat state generated form HHG we consider the problem of phase estimation. The setup under consideration is a beam-splitter parameterized by the angle $\theta$~\cite{campos1989quantum}, and described via the unitary operation $\hat{U}_{\mathrm{BS}}(\theta) = e^{-i\theta\hat{H}}$, with BS Hamiltonian $ \hat H =- i (\hat{a}_0^{\dagger}\hat{a}_1 - \hat{a}_0\hat{a}_1^{\dagger})$. In particular, we consider a homodyne configuration in which the input signal state $\hat{\rho}_0$ is overlapped with a local oscillator (LO) in a coherent state $\ket{\beta_1}$.
The output state after the BS is then given by $\hat{\rho}_{\mathrm{out}} = \hat{U}_{\mathrm{BS}}(\theta)\hat{\rho}_{\mathrm{in}}\hat{U}^{\dagger}_{\mathrm{BS}}(\theta)$, where $\hat{\rho}_{\mathrm{in}} = \hat{\rho}_0 \otimes \dyad{\beta_1}$. 
To quantify the metrological sensitivity we use the quantum Fisher information (QFI), which for the single parameter BS and an arbitrary signal state $\hat{\rho}_0$ can be written as~\cite{yu2018maximal,MullerRigat_2023}
\begin{equation}
\begin{split}
\mathcal{F} = & \sum_{i=1}^{d}4\lambda_i\bra{\lambda_i}\hat{H}^2\ket{\lambda_i} \\ & - \sum_{i,j = 1}^{d}\frac{8\lambda_i\lambda_j}{\lambda_i+\lambda_j}|\bra{\lambda_i}\hat{H}\ket{\lambda_j}|^2,
    \label{eq:QFIdensitymatrix}
\end{split}
\end{equation}
where $\hat{\rho}_{\mathrm{in}} = \sum_{i=1}^d \lambda_i \dyad{\lambda_i}$ represents the spectral decomposition of the density matrix of the input state, while $d$ is the dimension of the matrix support, $\lambda_i$ are the eigenvalues, and $\ket{\lambda_i}$ are the eigenvectors. 
For the following analysis of the QFI for the noisy cat states, we want to emphasize the role of the relative phase between the signal state $\hat \rho_0$ and the LO coherent state $\ket{\beta_1}$. We take into account the relative phase by assuming real valued amplitudes for the cat states, while the LO amplitude is given by $\beta_1 =\abs{\beta_1} e^{i \chi}$ with relative phase $\chi$.
We further note that in optical experiments, the relative phase of two states is rarely known; thus, we follow the path of increasing the QFI without the knowledge of it. Surprisingly, the metrological power of the HHG cat state will turn out to be independent of this phase, giving it a more robust meaning than for the even/odd cat states.

\subsection{\label{sec:noise_robust}Metrological robustness of QFI}

\begin{figure}[ht]
    \centering
    \includegraphics[width=1\columnwidth]{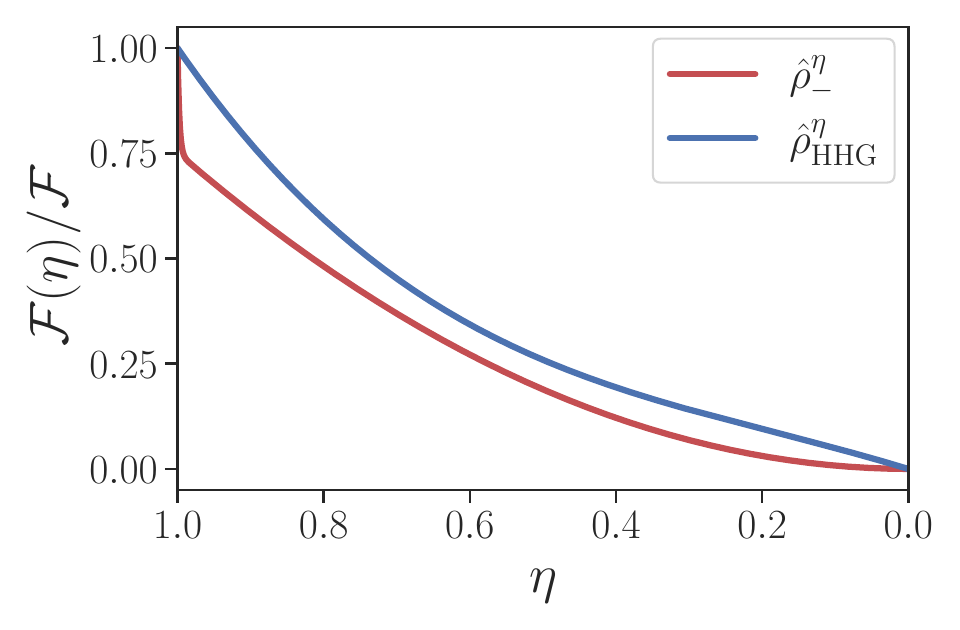}
    \caption{Robustness of the HHG-cat state compared to the odd cat state for varying losses. The noisy QFI $\mathcal{F}(\eta)$ are normalized with respect to the pure counterparts $\mathcal{F}$. We consider cat states in the high-photon number regime $\expval*{\hat N} = 100$ with a relative phase to the LO of $\chi = \pi /2$. The parameters used are $\alpha = 10$ for the odd cat state, and $\alpha' = 10.5$ and $\delta\alpha = -0.5$ for the HHG-cat. For the LO we have $|\beta| = 10$ (where $\beta = |\beta|e^{i\chi}$ for the pure case LO and $\beta = \sqrt{\eta}|\beta|e^{i\chi}$ for the reduced LO amplitude). }
    \label{fig:QFIlossynormalized}
\end{figure}

We shall now analyze if the enhanced robustness against decoherence due to photon loss of the HHG-cat state carries over to the problem of phase estimation leading to a metrological robustness.
To do so the QFI is used as the quantifier for the metrological power of the cat states for the phase estimation setup.
To calculate the QFI for the noisy cat states, we use the general expression, given by Eq.~\eqref{eq:QFIdensitymatrix}, and the eigenvalues from the spectral decomposition given in  Eq.~\eqref{eq:lossyevenodd} and Eq.~\eqref{eq:lossyHHGcat}.
The QFI for the homodyne phase sensing scheme using these noisy states is shown in Fig.~\ref{fig:QFIlossynormalized} for varying $\eta$, normalized to the QFI for the respective pure counterparts $\mathcal{F}(\eta) /\mathcal{F}$, and for an average photon number of $\expval*{\hat{N}}=100$ before losses. 

We can clearly see that the noisy HHG-cat state outperforms the noisy odd cat state in terms of the loss-induced relative change to the QFI. This holds for all values of photon loss, and the HHG-cat is particularly more robust in the regime of small losses. This regime is of notable interest since the photon losses are in general small, although not vanishing. For instance, an average photon loss of just a single photon for $\eta = 0.99$ (from the initial 100 photons) will give $\mathcal{F}_{HHG}(\eta) / \mathcal{F}_{HHG} = 0.975$ for the HHG cat, in contrast the odd cat state have $\mathcal{F}_{-}(\eta) / \mathcal{F}_{-} = 0.741$.
While the HHG-cat can protect most of its metrological sensitivity, the odd cat state does show a significantly reduced QFI due to losses. 
We note that for the relative phase $\chi = \pi / 2$ considered in Fig. \ref{fig:QFIlossynormalized} the odd cat state outperforms the HHG-cat in the pure case, while in contrast, the the HHG-cat shows its robustness against losses. 

\begin{figure}[ht]
    \centering
    \includegraphics[width=1\columnwidth]{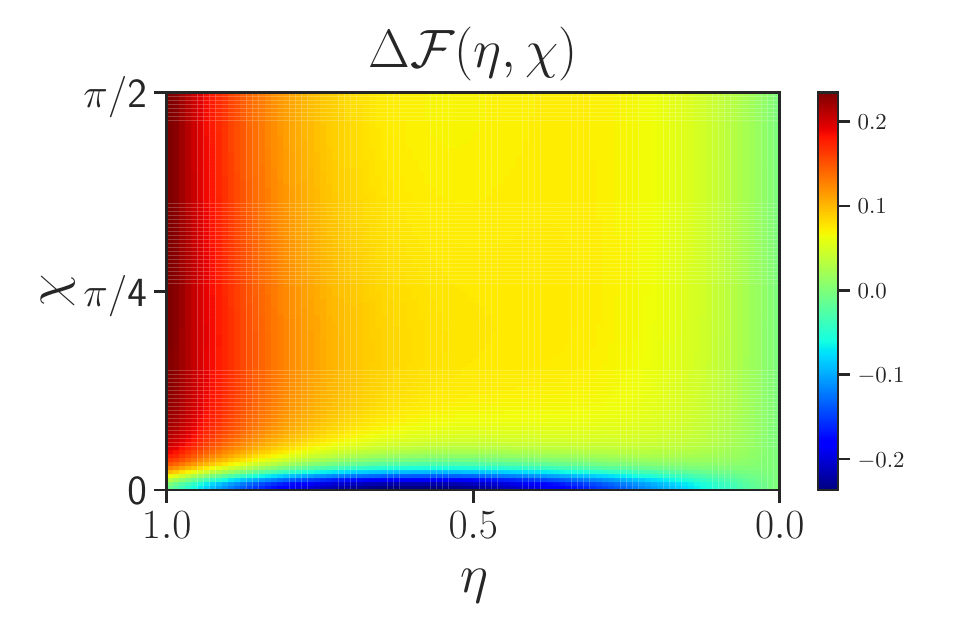}
    \caption{ Difference of the relative QFI changes for the noisy case $\Delta \mathcal{F}(\eta, \chi)$ for varying photon loss $\eta$ and LO phase $\chi$. For values $\Delta \mathcal{F} > 0$ the HHG-cat shows an enhanced robustness over the odd cat state. The photon number before losses is $\expval*{\hat N} = 100$, with the same parameters as in Fig.~\ref{fig:QFIlossynormalized} for the cat state amplitudes.}
    \label{fig:color_plot}
\end{figure}

However, so far in Fig.~\ref{fig:QFIlossynormalized} we have only considered the relative change of the QFI for a specific relative phase $\chi$ between the cat state and the LO. Therefore, in Fig.\ref{fig:color_plot}, we show the difference between the relative changes of the noisy QFI with its pure counterparts 
%\begin{align}
%    \Delta \mathcal{F}(\eta, \chi) = \mathcal{F}_{HHG} (\eta) / \mathcal{F}_{HHG} - \mathcal{F}_{-}(\eta) / \mathcal{F}_{-}.
%\end{align}
\begin{align}
    \Delta \mathcal{F}(\eta, \chi) = \frac{ \mathcal{F}_{HHG} (\eta) }{ \mathcal{F}_{HHG} } - \frac{ \mathcal{F}_{-}(\eta) }{ \mathcal{F}_{-} }.
\end{align}

If $\Delta \mathcal{F}(\eta, \chi)$ is positive for given $\eta$ and $\chi$ then the HHG-cat states are more robust to noise than the odd cat state. Conversely, the negativity of this notion shows higher usefulness of the odd cat state.
We observe a more robust HHG-cat state for the majority of values of the relative phase $\chi$, and especially a significant outperforming in the regime of small losses. 
Only for small values around the zero phase $\chi = 0$, the QFI of the odd cat state is less affected by photon loss. 
Interestingly, this is the regime in which the QFI of the pure HHG-cat is larger than the QFI of the odd cat state. 
This highlights the fact that the case in which the pure state has a larger QFI, the noisy state is more affected.

Finally, we observe that the noise range in which the HHG-cat state outperforms the even/odd cat states remains almost constant over four orders of magnitude of the average photon number. 
This shows that the metrological potential of the HHG-cat state is not restricted to a particular number of photons in the field, and is in agreement with the intuition that a higher QFI is more affected by the noise.

\subsection{\label{sec:phase_insensitivity}HHG-cat insensitivity to phase uncertainty}

Besides the inevitable occurrence of photon losses, for which we have seen the robustness of the HHG-cat over the odd cat state, the experimental parameters are never perfectly controlled. 
This also includes the relative phase between the probe state and the LO, which can fluctuate or simply be unknown, as in the majority of setups where the path difference is not precisely determined relative to the optical wavelength. 
In the following, we show that the HHG-cat state is metrologically robust against variations of this relative phase $\chi$, while the odd-cat state can exhibit significant deviations of the QFI for small changes in $\chi$.
To this end we compute the derivative of the QFI with respect to the relative phase $\partial_\chi [\mathcal{F}(\eta)/\mathcal{F}]$ in the case of small losses (solid lines for $\eta = 0.99$ and dashed lines for $\eta = 0.9$).
First, we can see that the derivative of the QFI for the HHG-cat vanishes for the entire range of relative phase $\chi$. This insensitivity with respect to the relative phase makes the HHG-cat very practical in the case where the phase is hard to control. 
In contrast, the odd cat state exhibits a strong dependence of the relative phase and can particularly change dramatically for small variations of the relative phase in the regime in which it is more robust than the HHG cat state. Thus, in the small region in which the odd cat state is more robust against losses over the HHG-cat, the QFI is very susceptible for changes of the relative phase with the LO while the HHG-cat preserves its QFI. 
This insensitivity of the HHG-cat with respect to the LO phase imperfections and the enhanced robustness against losses makes them a promising candidate for noise robust quantum metrology.

\begin{figure}[ht]
    \centering
    \includegraphics[width=1\columnwidth]{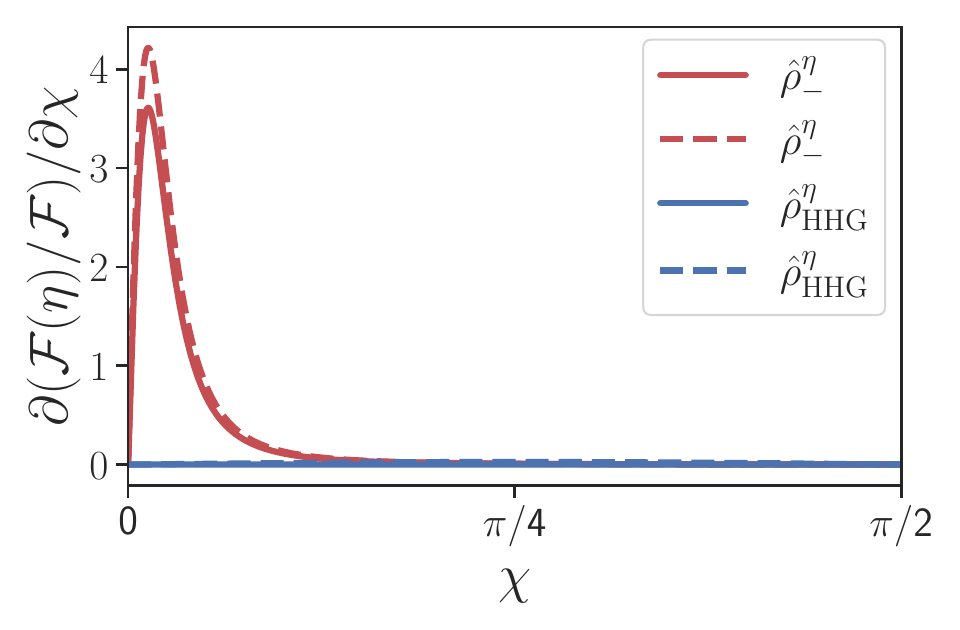}
    \caption{ Derivative of the relative QFI change with respect to the LO phase, i.e. $\partial_\chi [\mathcal{F}(\eta)/\mathcal{F}]$, for the HHG-cat (blue lines) and the odd cat state (red lines). The insensitivity of the QFI for the HHG-cat is exemplified for two values of the losses, $\eta = 0.99$ (solid) and $\eta = 0.9$ (dashed). The initial photon number before losses was set to $\expval*{\hat N} = 100$ with the same parameters as in the previous figures for the cat state amplitudes. }
    \label{fig:derivative}
\end{figure}

\subsection{The ideal scenario}

We have shown in Sec.\ref{sec:noise_robust} that the HHG-cat is more robust to noise and in Sec.\ref{sec:phase_insensitivity} that it is furthermore robust to phase changes. 
However, in Fig.\ref{fig:color_plot} we have seen that there exists a case in which the odd cat state is more robust than the HHG-cat, which is most pronounced for the relative phase $\chi = 0$.
We thus finally show, that in this case the HHG-cat is truly better than the traditional odd cat state in terms of its metrological usefulness in the ideal scenario without photon losses. 
Therefore, we consider the pure states given in Eq.\eqref{eq:cat_HHG} and Eq.\eqref{eq:cat_conventional} for the HHG-cat and the odd cat state, respectively. Considering that these states are pure $\hat \rho_{\mathrm{in}} = \dyad{\Psi_{\mathrm{in}}}$, the QFI in \eqref{eq:QFIdensitymatrix} simplifies to 
\begin{align}
    \mathcal{F} = 4 \left[ \bra{\Psi_{\mathrm{in}}} \hat H^2 \ket{\Psi_{\mathrm{in}}} - \bra{\Psi_{\mathrm{in}}} \hat H \ket{\Psi_{\mathrm{in}}} \right].
\end{align}

\begin{figure}[ht]
    \centering
    \includegraphics[width=1\columnwidth]{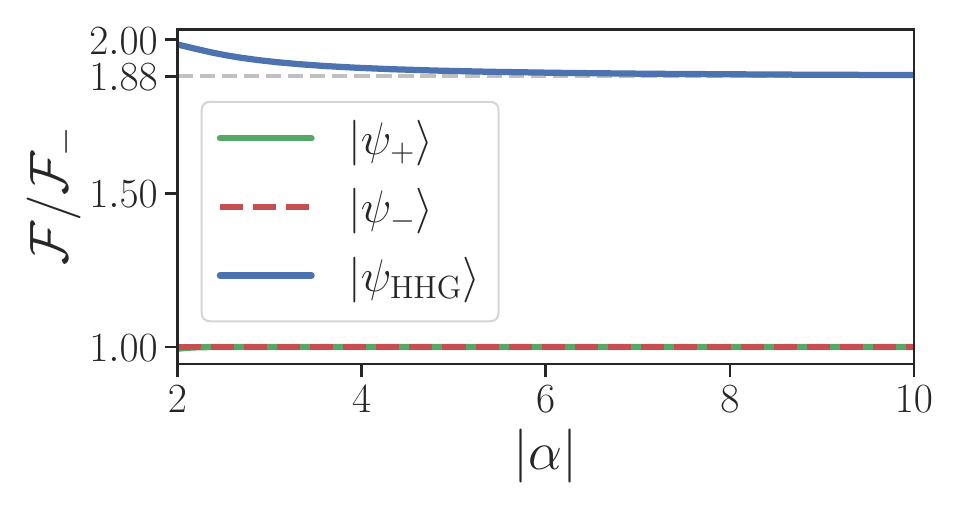}
    \caption{ QFI for the phase sensing setup using different pure signal states overlapped with a coherent state LO, normalized to the QFI for a pure odd cat state ($\mathcal{F}/\mathcal{F}_{-}$). The coherent state amplitude of the even/odd cat states is given by $\abs{\alpha}$, and for the HHG-cat state we have $\alpha = \alpha' + \delta\alpha$ for a fixed displacement $\delta \alpha = -0.5$, such that $\alpha' \in [2.5, 10.5]$ (therefore the mean photon numbers of all three corresponding input states only differ by less than a single photon). The mean photon number of the LO is set to match the mean photon number of the input cat states, $|\beta_1|^2 = |\alpha|^2$. For high values of the mean photon number, the QFI for the HHG-cat state input converges towards $\mathcal{F}_{\mathrm{HHG}} \approx 1.88\times\mathcal{F}_\pm$. }
    \label{fig:QFI_pure}
\end{figure}

The QFI for the pure cat states is shown in Fig.\ref{fig:QFI_pure} for increasing average photon number $\expval*{\hat N}$, and we observe that the HHG-cat state outperforms the even/odd counterparts in the entire region of high photon numbers $\expval*{\hat N} \ge 4$.
For small photon numbers the QFI of the HHG-cat is almost twice as much as for the traditional even/odd cat states, and as the coherent state amplitude increases, the enhancement of the QFI stabilizes at around $\mathcal{F}_{\mathrm{HHG}} \approx 1.88 \times \mathcal{F}_{\pm}$. 
We emphasize that this further manifests the metrological advantage of the HHG-cat over the traditional even/odd cat states in terms of the robustness to realistic scenarios including experimental imperfections.  

\section{Conclusions}

We have demonstrated that the optical cat states obtained from quantum state engineering approaches using the process of high harmonic generation~\cite{stammer2023quantum} show an enhanced robustness against photon losses compared to the traditional even/odd cat states in the regime of high photon numbers.
Especially, the HHG-cat is significantly more robust in the regime of small noise in which the loss of just a few photons from the initial 100 photons brings the even/odd cat state into a maximally mixed state while the HHG-cat remains an almost pure state.
Focusing on the realistic scenario in which losses occur and experimental imperfections are present, this translates to an enhanced robustness of the QFI for fields subjected to photon loss. 
Further, we show that the HHG-cat state is insensitive to the relative phase of the LO in the homodyne configuration, while the traditional even and odd cat states show distinct deviations in the QFI for changes in the relative phase.
In summary, the traditional even and odd cat states do not provide robustness to photon loss and phase uncertainty for metrological applications. 
In contrast, the HHG-cat state we have introduced for phase estimation achieves full robustness to phase changes as well as significantly outperforms the conventional cat states with respect to decoherence induced by photon loss. 

This is of particular importance for protecting quantum technologies against noise and experimental imperfections~\cite{suter2016colloquium}, and provides further insights into the applicability of the connection between attosecond physics with quantum optics and quantum information science~\cite{bhattacharya2023strong, lewenstein2022attosecond, cruz2024quantum}.

\begin{acknowledgments}

P.S. acknowledges funding from the European Union’s Horizon 2020 research and innovation programme under the Marie Skłodowska-Curie grant agreement No 847517. 
ICFO group acknowledges support from: ERC AdG NOQIA; MCIN/AEI (PGC2018-0910.13039/501100011033, CEX2019-000910-S/10.13039/501100011033, Plan National FIDEUA PID2019-106901GB-I00, Plan National STAMEENA PID2022-139099NB-I00 project funded by MCIN/AEI/10.13039/501100011033 and by the “European Union NextGenerationEU/PRTR" (PRTR-C17.I1), FPI); QUANTERA MAQS PCI2019-111828-2); QUANTERA DYNAMITE PCI2022-132919 (QuantERA II Programme co-funded by European Union’s Horizon 2020 program under Grant Agreement No 101017733), Ministry of Economic Affairs and Digital Transformation of the Spanish Government through the QUANTUM ENIA project call – Quantum Spain project, and by the European Union through the Recovery, Transformation, and Resilience Plan – NextGenerationEU within the framework of the Digital Spain 2026 Agenda; Fundació Cellex; Fundació Mir-Puig; Generalitat de Catalunya (European Social Fund FEDER and CERCA program, AGAUR Grant No. 2021 SGR 01452, QuantumCAT \ U16-011424, co-funded by ERDF Operational Program of Catalonia 2014-2020); Barcelona Supercomputing Center MareNostrum (FI-2023-1-0013); EU Quantum Flagship (PASQuanS2.1, 101113690); EU Horizon 2020 FET-OPEN OPTOlogic (Grant No 899794); EU Horizon Europe Program (Grant Agreement 101080086 — NeQST), ICFO Internal “QuantumGaudi” project; European Union’s Horizon 2020 program under the Marie Sklodowska-Curie grant agreement No 847648; “La Caixa” Junior Leaders fellowships, La Caixa” Foundation (ID 100010434): CF/BQ/PR23/11980043. Views and opinions expressed are, however, those of the author(s) only and do not necessarily reflect those of the European Union, European Commission, European Climate, Infrastructure and Environment Executive Agency (CINEA), or any other granting authority. Neither the European Union nor any granting authority can be held responsible for them.  

\end{acknowledgments}

\bibliographystyle{unsrt}
\bibliography{literatur}{}
%\end{multicols}

\appendix
\section{\label{sec:appendix_HHG}Eigendecomposition of the noisy HHG-cat state}

Here we show the exact expressions for the eigendecomposition of the noisy HHG-cat state subjected to losses from Sec. \ref{sec:noisy_cats}.
In Eq.\eqref{eq:lossyHHGcat} we have expressed the noisy HHG-cat state in its eigendecomposition 
\begin{equation}
    \hat{\rho}_{\mathrm{HHG}}^{\eta} = \lambda^{\eta}_{+}\dyad*{\lambda^{\eta}_{+}} + \lambda^{\eta}_{-}\dyad*{\lambda^{\eta}_{-}},
\end{equation}
where the eigenvalues are given by 
\begin{equation}
\begin{split}
\lambda^{\eta}_{\pm} = & \frac{1}{2N_{\mathrm{HHG}}} \Big[N_{\mathrm{HHG}} \pm \big[N_{\mathrm{HHG}}^2 \\ & - 4N^{\eta}_{\mathrm{HHG}}(|\xi^{\eta}|^2 - |\xi|^2 - |\xi^{\eta} - \xi\mu^{\eta}|^2)\big]^{1/2}\Big],
    \label{eq:lossyHHGeigenvalues}
\end{split}
\end{equation}
with $\xi^{\eta} = \braket{\alpha\sqrt{\eta}}{(\alpha+\delta\alpha)\sqrt{\eta}}$, $N^{\eta}_{\mathrm{HHG}} = 1 - |\xi^{\eta}|^2$, and $\mu^{\eta} = \braket*{(\alpha+\delta\alpha)\sqrt{1-\eta}}{\alpha\sqrt{1-\eta}}$, and with the corresponding eigenvectors
\begin{equation}
\begin{split}
\ket*{\lambda^{\eta}_{\pm}} = & \frac{1}{\sqrt{N^{\eta}_{\pm}}}\big[C_{\pm,1}^{\eta}\ket*{(\alpha + \delta\alpha)\sqrt{\eta}} \\ & + C_{\pm,2}^{\eta}\ket*{\alpha\sqrt{\eta}}\big],
    \label{eq:lossyHHGeigenvectors}
\end{split}
\end{equation}
where $N^{\eta}_{\pm}$ are normalization factors, and
\begin{equation}
\begin{split}    
C_{\pm,1}^{\eta} = & \frac{[(\xi^{\eta})^* - \xi^*(\mu^{\eta})^*]\sqrt{N^{\eta}_{\mathrm{HHG}}}}{N_{\mathrm{HHG}}\lambda^{\eta}_{\pm} - N^{\eta}_{\mathrm{HHG}}},\\
C_{\pm,2}^{\eta} = & \sqrt{N^{\eta}_{\mathrm{HHG}}} - \xi^{\eta} C_{\pm,1}^{\eta}.
\label{eq:lossyHHGeigenvectorsterms}
\end{split}
\end{equation}
% \begin{equation}
%     \begin{split}
%         \hat{\rho}_{\mathrm{HHG}}(\eta) = & \frac{1}{N_{\mathrm{HHG}}}(\ket*{(\alpha+\delta\alpha)\sqrt{\eta}}\bra*{(\alpha+\delta\alpha)\sqrt{\eta}} \\ & - \xi^*\mu^*\ket*{(\alpha+\delta\alpha)\sqrt{\eta}}\bra*{\alpha\sqrt{\eta}} \\
%         & - \xi\mu\ket*{\alpha\sqrt{\eta}}\bra*{(\alpha+\delta\alpha)\sqrt{\eta}} \\ & + |\xi|^2\ket*{\alpha\sqrt{\eta}}\bra*{\alpha\sqrt{\eta}}),
%     \end{split}
%     \label{eq:HHG}
% \end{equation}
%where $\mu = \braket*{(\alpha+\delta\alpha)\sqrt{1-\eta}}{\alpha\sqrt{1-\eta}}$ and $\xi = \braket{\alpha}{\alpha+\delta\alpha}$.

\section{\label{sec:appendix_derivative}Sensitivity to small photon loss}

\begin{figure}[ht]
    \centering
    \includegraphics[width=1\columnwidth]{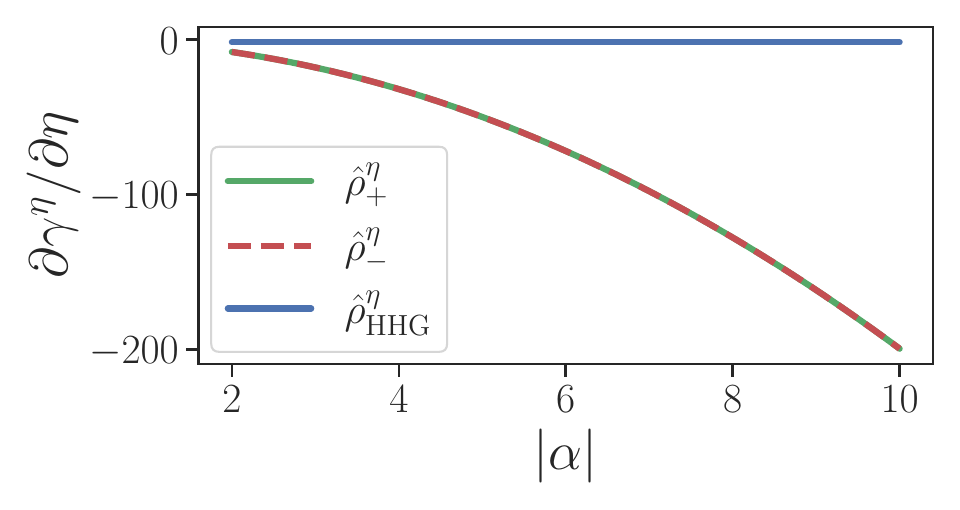}
    \includegraphics[width=1\columnwidth]{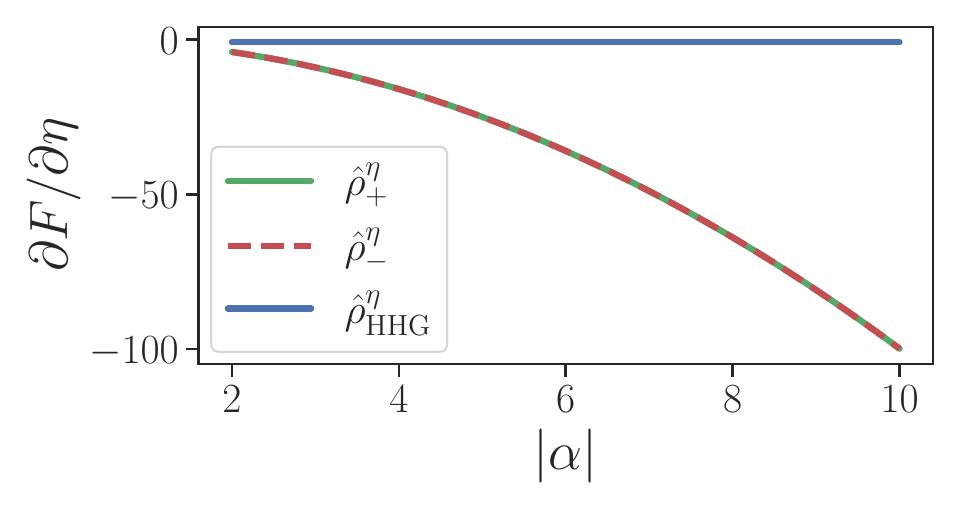}
    \caption{Derivatives $\partial_\eta $ of the purity $\gamma(\eta)$ in (a) and the fidelity $F(\eta)$ in (b) for the cat states subjected to photon loss evaluated at $\eta = 1$ for increasing average photon number $\expval*{\hat N}$. The amplitudes of the even/odd cat state range from $\alpha \in [2,10]$ and for the HHG-cat we have $\alpha \in [2.5, 10.5]$ for fixed $\delta \alpha = -0.5$, such that all cat states have approximately the same average photon number.  }
    \label{fig:fidelity_derivative}
\end{figure}

In Fig.~\ref{fig:fidelitynoisy} we have shown the purity and the fidelity of the noisy cat states for varying photon loss parameterized by $\eta$. In particular, the regime of small photon losses is interesting since it showcase that already tiny photon losses can have a tremendous influence on the optical state. 
Therefore, in Fig.~\ref{fig:fidelity_derivative} we show the sensitivity to small photon loss
\begin{align}
    \left. \pdv{\gamma(\eta)}{\eta} \right|_{\eta = 1}, \qquad \left. \pdv{F(\eta)}{\eta} \right|_{\eta = 1},
\end{align}
which is given by the derivative of the purity and the fidelity with respect to the photon loss parameter evaluated at $\eta = 1$. 
We can see that the HHG-cat state is insensitive to small photon loss for the entire range of average photon numbers, while in contrast the even and odd cat state show a reduced purity and fidelity. This loss in purity and overlap with the pure state significantly increases for increasing average photon number, which underlines the pronounced sensitivity to small photon loss.

\end{document}